# REVIEW SUMMARY

**COMPLEX SYSTEM ADAPTATION**

## Plasticity-rigidity cycles: A general adaptation mechanism

Peter Csermely*

**BACKGROUND:** Adaptation is essential for survival. System plasticity and rigidity were recurrently shown to be involved in adaptation mechanisms, and were described as i.) big and small phenotypes of human metabolism; ii.) initial and well-developed states of hematopoietic cell differentiation; iii.) more and less flexible modules human brains before and after accomplishing a learning task, as well as iv.) destabilized and re-stabilized states of memory reconsolidation, respectively. The stability-plasticity dilemma was considered as a basic constraint for both biological and artificial learning models.

Sequential simulated annealing, consisting of repeated heating and cooling steps, was shown to be a more efficient method to find the global optimum than single-round simulated annealing. Importantly, many other widespread search heuristics and optimization methods, such as Kalman-filtering, genetic algorithms, reinforcement learning or multiplicative weight updates, often use recursive cycles of less constrained, more plastic, discovery-type starting phase followed by a more rigid, selection-type optimization step.

**ADVANCES:** The efficiency of alternating plastic-like and rigid-like states in the adaptation to changing environments was shown in a large variety of complex systems. However, this extensive knowledge remained fragmented, and the generality of the phenomenon has not been described yet. Here plasticity-rigidity cycles are described as a general, powerful, system-level adaptive mechanism.

First, structural and functional plasticity and rigidity are both defined, where functional plasticity and rigidity describe many and few system attractors, respectively. Then, summarizing recent key publications, several mechanisms are listed, how plasticity-rigidity cycles optimize the system's structural stability. As a key point, I show the generality of plasticity-rigidity cycles reviewing the salient examples of simple molecular assemblies, assisted protein folding, cellular differentiation, learning, creative thinking and the efficient functioning of social groups and ecosystems.

The limitations of the concept, such as the complexity of multiple and overlapping plasticity-rigidity cycles in social- or ecosystems, is also summarized. Finally, important examples of the consequences of plasticity-rigidity cycle-induced adaptive processes are described, such as novel aspects of aging and evolvability, as well as the design of efficient therapeutic interventions in medicine and crisis management of financial and biological ecosystems.

**OUTLOOK:** Plasticity-rigidity cycles utilize and extend the duality of Archilochus' famous saying that "*The fox knows many things; the hedgehog one big thing*" elaborated by Isaiah Berlin. I hope that this work will prompt to develop more general numerical measures of both structural and functional plasticity/rigidity. Examination of plasticity-rigidity cycles i.) observed during environmental changes; ii.) working in multilayer networks of cells, brain, society or ecosystems or iii.) operating during evolution will also be exciting tasks of future studies. In an early study Dunham *et al.* described iterative improvement as the natural framework for heuristic design. Here I demonstrate that plasticity-rigidity cycles form a natural framework for adaptation in a wide range of complex organizations and life. ∎

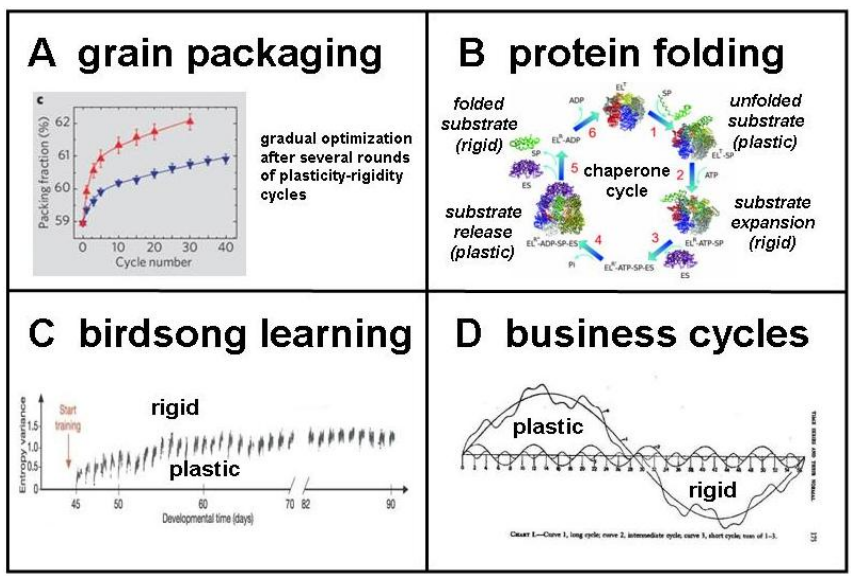

**Key examples of plasticity-rigidity cycle-mediated adaptation mechanisms.** The four illustrative examples cover a wide range of disciplines and system complexity. **(A)** Increase of packaging efficiency after thermally-induced changes in plasticity/rigidity of glass spheres in a plastic cylinder [reproduced with permissions from Chen et al. (*41*)]. **(B)** Molecular chaperones help protein folding *via* multiple rounds of extension and consequent release of substrates (*43*) inducing their larger rigidity and plasticity, respectively [reproduced with permissions from Stan et al. (*42*)]. **(C)** Cyclic development of song beauty (represented as Wiener entropy variance) in a 45-day training period of zebra finches. Birds that showed larger cycling in their song complexity learned better [reproduced with permissions from Derégnaoucourt et al. (*40*)]. Later studies extended these findings to human infants (*79*), and other bird species and showing that neuronal plasticity-rigidity cycles of the HVC bird song nucleus accompanied the sequential birdsong learning steps (*80*). **(D)** Business cycles described by Schumpeter (*103*) have several overlapping innovative/expansive phases and stagnating/selective phases. Business cycles increase overall productivity, and can be regarded as an adaptation process at the level of macro-economy. [Reproduced with permission from Schumpeter (*103*).] Organizational learning cycle (*95*) and changes of exploration and exploitation phases (*24*) describe other economy-related examples, where the operation of plasticity-rigidity cycles was more directly demonstrated.

*Semmelweis University, Department of Medical Chemistry, Budapest, Hungary
E-mail: csermely.peter@med.semmelweis-univ.hu

# Plasticity-rigidity cycles:
# A general adaptation mechanism

Peter Csermely[1,*]

**Successful adaptation helped the emergence of complexity. Alternating plastic- and rigid-like states were recurrently considered to play a role in adaptive processes. However, this extensive knowledge remained fragmented. In this paper I describe plasticity-rigidity cycles as a general adaptation mechanism operating in molecular assemblies, assisted protein folding, cellular differentiation, learning, memory formation, creative thinking, as well as the organization of social groups and ecosystems. Plasticity-rigidity cycles enable a novel understanding of aging, exploration/exploitation trade-off and evolvability, as well as help the design of efficient interventions in medicine and in crisis management of financial and biological ecosystems.**

**A**daptation is essential for survival. Moreover, series of successful adaptation steps helped the emergence of complexity contributing to the appearance of life, consciousness and human culture (*1*). Starting from the early conceptualization of homeostasis by Bernard and Cannon several adaptation mechanisms have been described, which either restore the original status or explore alternative options. Selye identified general mechanisms of adaptation to a sudden change of the environment, called stress. Waddington coined the term homeorhesis describing adaptive processes of dynamically changing systems reaching novel states. System-level adaptation mechanisms were described by Wiener, Ashby and von Bertalanffy. Degeneracy was shown as an efficient form of system-level adaptation increasing system robustness (*2-4*).

System plasticity and rigidity were recurrently considered to play a role in adaptive processes. Relatively separated states with more plastic and more rigid properties were described as big and small phenotypes of human metabolism (*5*); initial and well-developed states of hematopoietic cell differentiation (*6*); more and less flexible modules human brains before and after accomplishing a learning task (*7*), as well as destabilized and restabilized states of memory reconsolidation (*8*), respectively. The stability-plasticity dilemma (phrased as the bias-variance dilemma in stochastic learning) was considered as a basic constraint for both biological and artificial learning models (*9*). Sequential simulated annealing [also called as thermal cycling (*10*)] consisting of repeated heating and cooling steps, was shown to be a more efficient method to find the global optimum than single round simulated annealing (*11*). Importantly, a high number of widespread search heuristics and optimization methods, such as Kalman-filtering (*12*), genetic algorithms (*13*), reinforcement learning (*14*) or multiplicative weight updates (*15*) etc. start from a less constrained (more plastic, discovery-type) starting state followed by an optimization/selection step. These artificial intelligence optimization methods often include recursive cycles of the above two steps.

All in all, a wide array of methods and publications showed the efficiency of alternating plastic-like and rigid-like states in the adaptation to changing environments by a large variety of complex systems. However, this extensive knowledge remained fragmented, and the generality of the phenomenon has not been described yet. Here plasticity-rigidity cycles will be described as a general, powerful, system-level adaptive mechanism. (For definitions of plasticity and rigidity see Box 1, their numerical descriptions are described in Box 2.) Several mechanisms will be listed, how plasticity-rigidity cycles may find global optima at successive points of system development. As a key point of this contribution, I will show the generality of plasticity-rigidity cycles in the adaptation of a wide range of systems including the organization of simple molecular assemblies, as well as assisted protein folding, cellular differentiation, learning, creative thinking and the efficient functioning of social groups and ecosystems. The limitations of the concept, such as the current lack of a general mathematical framework connecting structural and functional plasticity/rigidity or the complexity of multiple, overlapping plasticity-rigidity cycles in social or ecosystems, will also be summarized. Finally, important examples of the consequences of plasticity-rigidity cycle-induced adaptive processes will be described, such as a novel understanding of aging and evolvability, as well as the design of efficient therapeutic interventions in medicine and crisis management of financial and ecosystems.

---

[1] Semmelweis University, Department of Medical Chemistry, H-1444 Budapest 8, Hungary
***Corresponding author.** E-mail: Csermely.Peter@med.semmelweis-univ.hu

## Definition and properties of plasticity and rigidity of complex systems

***Plasticity and rigidity are widely used, but ill-defined concepts.*** Plasticity describes non-reversible deformations in material science (*16*). Changes of synaptic plasticity constitute a major learning mechanism (*17*). As the third and last illustrative example for the wide-spread use of the term "plasticity", phenotypic plasticity plays a major role in evolution (*18*). The first mathematical description of structural rigidity was given by Maxwell (*19*). The connectivity-dependent minimal condition for the rigidity of 2-dimensional networks of rods and joints was discovered almost a hundred years later by Laman (*20*). However, the 3-dimensional extension of Laman's theorem is still an important open problem of combinatorial rigidity in mathematics. Behavioral rigidity was re-defined several times since the end of 19$^{th}$ century (*21*). Plasticity and rigidity are related to the major concepts of system stability (*22,23*), complexity (*1,4,23*), degeneracy, robustness (*2*), evolvability/canalization (*18*), the exploration/exploitation trade-off (*24*), etc. These are only a few examples of the widespread use of plasticity and rigidity in various disciplines, which illustrate that plasticity and rigidity are important, but rather hidden hubs of human conceptual thinking (Box 1 and Fig. 1A).

***Controversial aspects of plasticity and rigidity: robustness, flexibility and stability.*** Definitions of plasticity and rigidity often focus on different segments of the related network of key concepts (Fig. 1A). Moreover, from the system's point of view plasticity and rigidity are both emergent properties, which can only be observed at the system-level. Due to this complexity the relationship of plasticity and rigidity to a number of other concepts, such as robustness, flexibility or stability, remained rather controversial. Formerly, plasticity was often considered as the opposite of robustness. However, later system plasticity was shown as an important requirement of robustness (*2,25*). Flexibility is still often used as a synonym of plasticity. However, flexibility is related to fast and reversible system responses to well-defined changes in the environment, which is more characteristic to (partially) rigid than plastic systems (Box 1 and Table S1 of Supplementary Information).

Controversy around the concepts of plasticity and rigidity is centered on the various concepts of system stability (Box 1). Lyapunov stability describes the speed, how a simple system moves back to its equilibrium state after a perturbation. Relatively simple plastic and rigid systems display a low and high Lyapunov stability, respectively (Fig. 1B and Table S1 of Supplementary Information). (Note that Lyapunov stability can be extended to stochastic conditions, see Box 1.) More complex systems have a state space with more than one 'local equilibrium condition', i.e. attractor. Importantly, starting with the pioneering work of Kauffman, several studies showed that the number of attractors of real world complex systems is surprisingly low (*26-28*). Changes of the system or its environment change the system's attractor structure. Structural stability (*22*) implies that the system's attractor structure remains rather similar withstanding a wider range of system or environmental changes. Several studies implied or showed different changes of Lyapunov and structural stabilities in various scientific disciplines, such as cellular functions, cancer progression, morphological development, ecosystem stability, etc. (*23,26,29*). However, the generalization of the differences between Lyapunov stability and structural stability is missing, which is the major cause of the controversy around plasticity, rigidity and related concepts.

Lyapunov stability of complex systems depends on the structure of their actual attractor. Plastic and rigid systems have smooth and rough state spaces, with typically low and high Lyapunov stabilities, respectively. Importantly, neither extremely plastic, nor extremely rigid systems display a high structural stability. Extremely plastic, 'fluid' systems change their attractor structure easily. Extremely rigid systems are fragile, which may lead to their decreased structural stability inducing large gross changes in their attractor structure, once the environmental changes go beyond their limited response set, which was acquired and refined by the rigid systems during their past experiences. Structural stability is maximal in complex systems displaying signs of both plasticity and rigidity (Fig. 1B). Systems displaying high structural stability are rare, but (fortunately) can be accessed easily by relatively few changes of suboptimal systems (*30*). Indeed, increased structural stability is required for fast adaptation, if the attractor structure is complex, i.e. not all attractors are directly accessible from any other attractors [which is true for state spaces of all complex systems (*25*)].

Flexibility (Box 1) is a property of rigid systems implying a high Lyapunov stability. Robustness is related to high structural stability, and thus, it requires both system rigidity and plasticity. On one hand, a certain extent of robust behavior may be displayed by rigid systems having e.g. negative feedbacks. On the other hand, robustness to unpredictable changes of the environment is a property of (partially) plastic systems (*2,31*), and is characterized often by low Lyapunov, but high structural stabilities. The low Lyapunov stability of plastic systems was the primary reason, why originally robustness was considered as the opposite of plasticity.

***(Re)definition of plasticity and rigidity.*** Plasticity and rigidity are defined in this paper as opposite characters of system's responses. A plastic system has a high number of possible responses to environmental changes. On the contrary, a rigid system may display only a low number of possible responses upon a similar stimulus. (Box 1). Responses of most complex systems are probabilistic, and their probability may widely differ. Thus, the "number of possible responses" may be better approximated by the entropy of the system's transition probability distribution (Box 2). Since these definitions are related to the function of the whole system, they define functional plasticity and rigidity. Structural plasticity and rigidity are defined as structural properties of the networks describing complex systems (mathematically most specified to networks of rods and joints, see Box 2). Functional and structural plasticity and rigidity are not synonym terms (*23,26,29*), though there is a large overlap between functionally and structurally plastic/rigid systems.

***Properties of plastic and rigid systems.*** Complex systems may often display a bimodal distribution, where one of the two abundant system states has a greater plasticity, while the other has a larger rigidity. Simple systems may display extreme forms of plasticity or rigidity. Extremely plastic systems explore a large number of possible solutions, and thus, are highly adaptive to (even) unexpected changes of their environment. However, these extremely plastic systems do not have a 'memory' (Box 1), thus they can not reliably and efficiently produce the same optimal response to a repeated stimulus. Extremely plastic systems dissipate perturbations well, and thus transmit them poorly. On the contrary, extremely rigid systems are highly optimized to a rather limited set of responses, are able to reproduce these responses reliably, but can not adapt to unexpected changes of their environment. Extremely rigid systems transmit perturbations well, and dissipate them poorly [for a summary of plastic and rigid system properties, see Table S1 of Supplementary Information (*31,32*)].

**Plasticity-rigidity cycles as a general adaptive mechanism**
Complex systems never display extreme forms of plasticity or rigidity permanently, but change these properties dynamically. When complex systems are shifted from a more plastic to a more rigid state, their response-set becomes limited, and their optimal (or quasi-optimal) responses become selected. When complex systems are shifted from a more rigid state to a more plastic state, their response-set becomes wider exploring novel ('creative') options of adaptation (*31*).

Complex systems often cycle between plastic and rigid states repeatedly. I term these cycles as "plasticity-rigidity cycles". Plasticity-rigidity cycles help the system to find its global optimum (Fig. 1B), similarly to that in many artificial intelligence methods, such as colored-noise recursive Kalman-filtering (*12*), adaptive genetic algorithms (*13*), variable discount factor reinforcement learning (*14*), learning and re-learning in Boltzmann machines, random dropout neural networks (*33*), weight redistributive multiplicative weight updates (*15*), or sequential simulated annealing (*10*) all using alternating exploration/discovery-type and optimization/selection steps. The universality of the dual concept of chance and necessity has already been coined by Democritus, and was beautifully expanded by Jacques Monod (*34*). A number of recent studies showed that discovering exploration and optimized experience are key factors of successful adaptation of cells, organisms, animal and human communities (*35-37*).

*I define plasticity-rigidity cycles as alternating changes of plasticity- and rigidity-dominance in system behavior resetting the ratio of plastic and rigid system properties to find the maximal structural stability of the system in the changed environment.* As I will demonstrate by a wide range of salient examples later, plasticity-rigidity cycles emerge as a powerful, general, system-level adaptive mechanism in all areas of natural organization and life.

Plasticity-rigidity cycles proceed *via* (relatively) smooth → rough → smooth state spaces as the system changes from a more plastic to a more rigid, and consecutively, to a more plastic state again. The repeated changes of state space roughness provide repeated 'windows of opportunity', where the system can be easily re-programmed passing a bifurcation *via* a critical transition (*38*).

The relative length and intensity of plasticity and rigidity changes can be widely varied. However, it is of key importance, that system constraints should prevent both plasticity and rigidity 'overshoots' during plasticity-rigidity cycles. Mechanisms of this safeguarding behavior have not been elucidated yet. However, system resource limits and stochastic node behavior help to prevent the development of too much system plasticity and rigidity, respectively. Quenched thermal cycling with slowly decreasing changes was proved to be rather efficient to find the global optimum in relatively simple cases (*10*). A cycle-asymmetry having longer rigid than plastic phases may reflect well conditions, where abundant system resources (inducing more plastic conditions) are depleted fast by a growing and competing population of systems. Longer rigid than plastic phases may be necessary due to the larger difficulties in system rearrangements after the completion of a rigidity transition.

Multiple (10 to 50) plasticity-rigidity cycles often occur in a wide range of natural optimization processes (*39-43*). Further studies are needed for the complete elucidation of the plasticity-rigidity cycle properties, which are most adequate to find the optimal response to different environmental changes.

## Mechanisms of plasticity-rigidity cycles

In this section the mechanisms are described, which help to induce and regulate plasticity-rigidity cycles. First, I will show how noise and medium effects (as 'network-independent' ways) modulate plasticity and rigidity. Next, I will summarize the current knowledge, how network structure may help the development of plasticity-rigidity cycles (see Table S2 of Supplementary Information for a summary).

*Network-independent mechanisms: noise*. Noise reduces the accuracy of signal transduction (where 'signals' were encoded by the system earlier as preferred ways of communication; Box 1). Noise may be induced by unexpected, stochastic changes of the system's environment (extrinsic noise), or by an increased internal plasticity of the nodes forming the network, which describes the system (intrinsic noise). As an example of this 'hierarchical plasticity propagation' an increased plasticity of individual neurons and their synapses increases the plasticity of the brain they form. Importantly, a low amount of properly positioned (e.g. inter-modular) 'noisy' nodes is enough to induce an increased plasticity at the system-level (*44,45*), as I will show in the next section on network-related mechanisms in detail.

Noisy systems generally have a high plasticity, while minimized noise results in a more rigid behavior. Importantly, in a large variety of complex systems increased noise helps to reach attractors, which were otherwise segregated by too large barriers in the system's state space (*27*). Noise-induced access of 'hidden' attractors can be used for the activation of latent diseases for their efficient therapy (*46*). Fluctuating noise induces plasticity-fluctuations, thus the increased efficiency of noise-assisted adaptation may actually derive from the induction of ('micro-scale') plasticity-rigidity cycles. In agreement with this, the combination of noise and long-term memory leads to the efficient development of cooperation independently of system configuration in spatial social dilemma games (*47*).

*Network-independent mechanisms: medium-effects*. Nodes of complex, real world networks are often embedded in a medium, i.e. in an environment, which may induce, or inhibit their plastic behaviour. Medium-effects primarily enable or disable nodes to make network contacts, thus 'melt' or 'freeze' the current network structure. A major mechanism of medium-effects is an increase or decrease of available system's resources. As I will show in the next section, system resource changes are driving forces of network topological phase transitions. Thus beyond a changing 'viscosity' of the system's environment, medium-effects may also directly influence the system's network structure.

Water-mediated lubrication, which is a primary requirement and regulator of the plasticity of protein structures, is an excellent example for the above medium-effects. Fluctuating intra-protein water content helps the induction of plasticity-rigidity cycles, and may play an important role in the mechanism of chaperonin-assisted optimization of unfolded or misfolded protein structures (*48*).

Chaperones have a high abundance, and display low-affinity interactions with a high number of other proteins. These properties pose chaperones as a 'medium' for a large subset of cellular proteins. Indeed, chaperones, like Hsp90, were shown to act as 'buffers' of evolvability. Here environmental stress leads to more misfolded proteins, which reduces active chaperone capacity leading to the appearance of previously buffered genetic variations at the level of the phenotype (*2,49*). Stress-mediated chaperone buffering acts like a plasticity-rigidity cycle, where low-stress/buffered phases correspond to plastic, while high-stress/non-buffered phases to rigid cycle segments. System-wide studies indicated that chaperone-induced evolvability-buffering may be a network effect displayed by many other proteins (*50*). In agreement with this, recent studies indicate a chaperone-like medium-effect of a number of prion- and amyloid-like proteins (*51*). Low affinity interactions ('weak links') were shown to increase the structural stability of complex systems from molecules to societies (*3*).

Recently, several novel mechanisms were suggested to regulate the fluidity of intracellular compartments, such as bacterial metabolic activity (*52*), myosin-related active random stirring of COS-7 cell cytoplasm (*53*) or fluctuating water-permeability of aquaporin plasma membrane channels (*54*). These mechanisms may help the development of plasticity-rigidity cycles regulating cell differentiation, cancer progression, or changes of synaptic plasticity. Membrane-remodeling was shown to induce higher plasticity of internal membranes than that of the plasma membrane after heat stress (*55*). Membrane fluidity changes may also modulate plasticity-rigidity cycles. Several forms of neuromodulation, such as volume transmission (*56*), may be regarded as a medium-effect at the neuronal network level.

In the society medium-effects are mediated by several 'system resources', such as trust, sharing of novel information in form of innovation and, most importantly, money (*57,58*). Janos Kornai in his seminal contributions described the two major social systems of the second half of the 20$^{th}$ century, capitalism and socialism, as surplus and shortage economies having high and low levels of innovation, respectively (*58*). Importantly, the surplus of system resources characterizes the more plastic, capitalist economy, while the shortage of system resources contributed to the rigidity of socialism.

***Network-related mechanisms: nodes and hubs.*** The development of plasticity-rigidity cycles can be initiated already at the level of individual network nodes. Node diversity, in forms of different internal node structure (note that in most real world networks, nodes are networks by themselves, Box 1), different levels of internal nodal noise, nodal plasticity and nodal rigidity, already play an important role in shifting the emergent plasticity and/or rigidity of the whole system, thus helping the system's optimization process (*44*). Plasticity (softening) and rigidity transitions may be initiated by special 'soft spot', or 'rigidity-seed' nodes, respectively. Soft spots may break rigid network structures similarly to lattice defects. Rigidity-seed nodes may establish rigid clusters (such as triangles or loops), and rigidity-promoting nodes may help the growth of these initial 'rigidity templates' causing a rigidity phase transition (*59,60*). Pre-stressed edges (*61*) may initiate these processes. Creative nodes (which are highly dynamic nodes bridging many network modules) are especially well-positioned to play a major role inducing plasticity/rigidity transitions (*45*).

As a rule of thumb, dense networks (with a large connectivity and/or high-weight edges) are usually more rigid than sparse networks. (Importantly, small networks are often denser, thus more rigid, than large ones. Large networks usually have more degrees of freedom (*62*). However, complex systems show obvious exceptions of these 'general rules'.) Edge density-induced rigidity poses hubs (i.e. nodes with a high number of neighbors, Box 1) as local rigidity centers. Party hubs, which have a constant neighborhood, are especially prone to increase network rigidity. Decrease of system resources (or increase in environmental stress) induce the emergence of hubs from a random network followed by the rearrangement of hub-structure developing a star-network having mega-hub(s), which attract most connections (*3*). This also increases network rigidity, similarly to the effect of increasing network hierarchy described later. Plastic and rigid phases of plasticity-rigidity cycles may display decreased and increased levels of hub-dominance, respectively.

***Network-related mechanisms: network core and modules.*** The network core is a densely connected central part of the network (Box 1). In directed networks this structure resembles to a bow-tie structure, where the core is in between of incoming and outgoing edges. Interconnected hubs form a rich-club, which is a widespread form of network cores. Large/fuzzy cores increase network degeneracy and system robustness, while small/compact cores make the network highly controllable, and induce non-reversibility (*2,4,31*). Decrease of system resources, or increase in environmental stress induce the development of smaller network cores. Importantly, hub-hub connections are often weak, since hubs' limited resources allow the maintenance only a few high intensity connections at the same time. This 'hub-repulsion' may increase core plasticity (*31*). Plasticity-rigidity cycles may involve a periodic increase and decrease of core size in plastic and rigid phases, respectively. Fluctuating 'hub-repulsion' may play an important role in plasticity-rigidity cycle development. Core structure in the form of nested networks was predicted to stabilize complex systems in the early work of May (*63*). Indeed, network nestedness was recently shown to increase the structural stability of ecosystems (*29*).

Network modules are densely connected groups of network nodes, which are relatively isolated from their environment (Box 1). Modularity was suggested to stabilize ecosystem networks already in 1972 (*63*). Temporal fluctuations of modular structure can be observed in networks of active brain neurons, which participate in plasticity-rigidity cycles related to learning processes (*7*). Modules become more condensed (and more isolated from each other) upon environmental changes including stress (*64*). Densely connected modules increase local rigidity, while the concomitant decrease in inter-modular contacts increases plasticity at the system-level. The development of network modules is important all the more, since they allow parallel optimization to multiple requirements at the same time. This might be an important reason, why varying environments induce modularity in molecular networks and in brain structure. Overlapping modules may operate in plastic, while compact modules may characterize rigid phases of plasticity-rigidity cycles. However, plasticity-rigidity cycles may also occur in different modules separately.

***Network-related mechanisms: network hierarchy and classes.*** Changes in network hierarchy form a key mechanism of plasticity-rigidity cycle development. Complex systems display two distinct control modes: distributed and centralized, which may correspond to the rigid and plastic phases of plasticity-rigidity cycles, respectively. Importantly, changes between distributed and centralized control may be induced by flipping the direction of a few, well-chosen network edges (*65*). Additionally, network hierarchy was shown to emerge as a response to changing environments, and induced a better performance than the sum of isolated nodes (*36*).

Flow type networks (like metabolic or signaling networks) often develop tighter cores than association-type networks, such as protein-protein interaction or social networks. The latter networks often display multi-modular structures (*31*). A recent study showed that source-dominated networks, such as neuronal and social networks, allow uncorrelated, relatively plastic behavior, while sink-dominated networks, such as transcriptional regulatory networks, display a hierarchical, correlated, relatively rigid behavior (*66*). Importantly, source-dominated and sink-dominated networks resemble the previously described surplus-dominated and shortage-dominated societies, capitalism and socialism (*58*), respectively. The plastic phase of plasticity-rigidity cycles may be more source-dominated, while the rigid phase may be more sink-dominated.

The above mentioned important differences between network classes may imply that sub-networks of the interdependent, multiplex networks of complex systems may display different levels of plasticity and rigidity at the same time (Table S2 of Supplementary

Information). As an example, metabolic networks may be more rigid than the related protein-protein interaction networks. The discrimination of network classes and their behavior is an exciting area of future studies. The contribution of node-structures (such as those described in this section: network cores, modules and hierarchy) and edge-structures [such as network skeletons or backbones (*31*)] to plasticity-rigidity cycles also awaits future clarification.

*Network-related mechanisms: fast and slow thinking of networks.* As described in a parallel publication in detail (*67*), recent data indicate that the adaptive response of many complex systems first mobilizes a fast, pre-set response of a well-connected network core. The fast-acting core provides responses pre-set by previous network experiences. If the core fails to reach a consensus, a majority of weakly linked, peripheral nodes generates novel responses. The consensus of the core dissipates the stimulus fast to the environment. If the stimulus is from a novel, unexpected situation, a conflict may develop between the stimulus and the constraints previously encoded by various segments of the rigid core. In such situations none of the encoded responses (appearing as attractors of the original network configuration) becomes stabilized, and the system fluctuates between these attractors remaining unstable. In this latter case the stimulus will propagate to the weakly connected, peripheral nodes, and a collective decision of (practically) the whole network, the "wisdom of crowds" stabilizes the system allowing a slow (but creative) dissipation of the stimulus. In case of a novel, unexpected stimulus, its slow dissipation may partially 'melt' the rigid core making the whole system more plastic. Increased plasticity helps to generate novel attractors or makes hidden attractors accessible. If the slow majority finds an optimal response/dissipation pattern, the same stimulus, if repeated, may modify the network (*via* Hebbian learning or similar processes) encoding a novel set of constraints. This makes the network more rigid again, reconfiguring its core ('electing new leaders') and enriching the system with a novel encoded response. Complex systems may adapt with an initial, approximate, but fast response of their core, which becomes refined by the inclusion of the periphery later. If none of these solutions work, the adequate response may be discovered by another complex system of a diverse population.

## Examples of plasticity-rigidity cycle-induced adaptation

In this section a number of examples are listed showing that plasticity-rigidity cycles can be identified as adaptive mechanisms of molecules, cells, human cognitive processes, social groups and ecosystems (Table S3 of Supplementary Information).

*Adaptation of molecular assemblies and granular materials.* The name of the plastic/rigid cycle(s) of the simulated annealing optimization method came from the wide-spread practice of annealing in metallurgy, where a heating/cooling cycle rearranges dislocations making the structure more homogeneous (*11*). Thermal cycles induce a tighter packing of granular materials, like glass or plastic spheres [Fig. 2A (*41*)]. The combination of physical strain and thermal cycles induces the growth of abnormally large grains of polycrystalline materials of shape-memory alloys resulting in the development of bamboo-type superelasticity opening potential seismic applications (*68*). Network rigidity was shown to induce cooperation of molecular structures leading to their self-organization (*69*). These examples show that the combined effects of heat-induced system plasticity and rigidity-induced cooperation may lead to the stabilization of either homogenous or heterogeneous phases depending on the 'life-history' of the system exemplified by a preceding strain (*68*).

*Adaptation of macromolecular structures.* Solving the Levinthal-paradox by finding the native state from the myriads of possible protein conformations is one of the most challenging optimization tasks in nature. Molecular chaperones actively help this protein folding optimization process by an ATP hydrolysis-driven chaperone cycle. Iterative annealing (i.e. repeated, ATP-dependent cycles of substrate binding and release; Fig. 2B) was reported as a general mechanism of the 60 kDa molecular chaperones and RNA-chaperones helping protein folding and complex formation. In a cycle of iterative annealing, chaperones first expand their substrate proteins, making them more rigid. Expansion is followed by substrate release making the substrate structure more unfolded, thus more plastic. As the substrate re-folds, its structure becomes more rigid again. These plasticity → rigidity → plasticity → rigidity transitions are repeated several times (*42,43,48*). Plasticity-rigidity changes also occur in the chaperone cycle of the 70 kDa chaperones, where extension and release of short peptide sequences (*39*) correspond to the plastic and rigid phases, respectively. Co-chaperone molecules often regulate the formation of more plastic or more rigid chaperone complexes, as exemplified by a recent study on the 90 kDa chaperone machine (*70*).

Plasticity-rigidity cycles may also operate at many other optimization processes of macromolecular conformations. Binding of substrates, allosteric modulators and, especially, other macromolecules generally increases the rigidity of local protein (or RNA) structures. However, the binding event (especially, if using the conformational selection mechanism) is a multi-step process (*31*), where transient plasticity increases may also occur. Importantly, transient increases and decreases in the size of disordered protein segments may also induce conformational plasticity-rigidity cycles helping to find the global optimum of macromolecular assemblies. The experimental proof for these complex, multimolecular plasticity-rigidity cycles is currently missing, and will be an important task of future studies.

*Cell differentiation and induction of pluripotent stem cells.* Stem cells have an exceptionally high plasticity, which enables their fast switches between the attractors of their smooth state space (*54*). Cell differentiation of progenitor cells proceeds *via* an initial increase in cellular plasticity followed by the development of more and more committed, rigid cellular networks, as it was shown by the entropy changes of gene co-regulation and chromosomal organization of differentiating hematopoietic progenitor cells. In the rigidity → plasticity → rigidity transitions of cell differentiation the end-state is *more* rigid than the starting state (*6*). Additionally, differentiation-induced epigenetic modifications make the state space rougher (*28*). The overall rigidity increase of differentiated cells is also displayed by the decrease of developmental plasticity (*5*), during biofilm formation of bacterial communities (*71*), as well as by the recent finding that priming of naïve T cells involves an initial stimulation of multiple cell fates followed by a selective, preferential expansion (*72*).

A seemingly reversed, rigidity → plasticity transition occurs during the induction of pluripotent stem cells. However, this process also has a later, maturation/stabilization phase, where the early increase in stochastic processes is followed by a more hierarchical phase of gene expression regulation (*73*). Thus both the induction and differentiation of stem cells displays a rigidity → plasticity → rigidity transition. However, in the induction of pluripotent stem cells the end-state is *less* rigid than the starting state (*73*).

In summary, cellular differentiation and dedifferentiation processes may be perceived as series of plasticity-rigidity cycles, where cells learn the requirements of the environment. An accelerated version of these adaptive cycles occurs in cancer development exemplified by the highly versatile plasticity-rigidity cycling of cancer stem cells (*26,54*) being more aggressive in recurrent than in primary tumors (*74*). Another key example of repeated plasticity-rigidity cycles is sexual reproduction, where epigenetic reprogramming proceeds *via* the erasure and re-establishment of epigenetic marks during early embryogenesis in the newly defined germ line and after fertilization in the zygote (*75*) inducing more plastic and rigid state spaces, respectively.

***Learning and memory formation.*** Synaptic plasticity was first described by Ramon y Cajal 120 years ago, was (re-)named by Konorski in 1948, and forms a major tenet of the Hebbian learning theory (*17*). Changes of synaptic plasticity play a key role in the formation of short- and long-term memories, in memory consolidation, reconsolidation and retrieval (*8*). Due to the large complexity of these overlapping processes as shown already by Bartlett describing schema development in 1932, the extent and direction of synaptic plasticity changes can not be predicted without the previous stimulation history. The ability to induce further changes in synaptic plasticity was termed as metaplasticity (*17*).

Several studies indicated that initially a high plasticity is required of successful encoding of the novel information, while memory formation induces a more rigid neuronal network structure. Such plasticity-rigidity cycles were shown in parvalbumin-interneuron networks during fear-conditioning (*76*); in rat medial prefrontal cortex interneuronal networks during the exploration of alternative strategies modifying previous beliefs (*77*); in mouse cortex networks during motor learning (*78*) and in human brain neuronal networks during motor skill training (*7*). Importantly, initial changes of neuronal plasticity (measured as the allegiance of active neurons to modules of other active neurons) were good predictors of the learning efficiency of human subjects in later training sessions (*7*).

A beautiful example of repeated learning cycles is shown on Fig. 2C, where the cyclic increase of the complexity of zebra finch song is shown during a training period of 45 days (*40*). Later studies extended these findings to other bird species and human infants (*79*). Neuronal plasticity-rigidity cycles of the HVC bird song nucleus accompanied the sequential birdsong learning steps (*80*). Repeated plasticity-rigidity cycles of serial mouse training sessions showed an initial expansion of the number of affected neurons followed by their gradual refinement into a smaller, targeted neuronal population (*78*). A highly similar pattern was observed in humans, where a high-intensity co-evolution of affected neuronal networks settled down with subsequent practice showing the emergence of an autonomous subnetwork (*81*). In agreement with these findings, memory reconsolidation was shown to be a gradually decreasing process, where upon more and more advanced memory consolidation less and less memory reconsolidation occurred (*17*).

The transiently increased neuronal plasticity during memory reconsolidation has very important applications. Fear conditioning can be inactivated and reactivated by the optogenic delivery of long-term depression and potentiation, respectively (*82*). Similar, optimally timed interventions were successfully used in the restoration of lost corticospinal tract connections after rat stroke (*83*), in treatment of post traumatic stress disorder, in increasing teaching efficiency, and in explaining suggestive questioning-induced witness memory distortions during criminal investigation (*17*).

Memory formation and memory reconsolidation are accompanied by intra-synaptic plasticity-rigidity cycles at the molecular level involving proteasome-dependent protein degradation and de novo protein synthesis in *Aplysia* (*8*), as well as a decrease and increase of actin cytoskeleton complexity in various organisms (*84*). Many network-independent and network dependent mechanisms, like changes in neuronal noise; effect of neuromodulators (including the effect of neuronal plasticity-increasing drugs, like Prozac); neuronal diversity; changes in hub, network core and hierarchy dominance; transient intermodular connections of spatially distributed brain regions, as well as formation of new neurons (*17,44,56,85*) play a key role in the regulation of plasticity-rigidity cycles at the neuronal network level. Plasticity-rigidity cycles displayed on multiple levels of brain architecture showing distinct kinetic patterns may significantly overlap with each other (*86*). A recent study showed that interconnections between neuronal and cerebral circulation networks induce the emergence of switch-type spontaneous synchronization (*87*). Interconnected brain networks may amplify plasticity-rigidity changes.

***Plasticity-rigidity cycles drive creativity.*** Increased human neuronal plasticity is characteristic to exploratory, creative periods (*85*). Positive emotions broaden the response repertoire (*88*), increase the plasticity of the brain's mindset, and boost creativity. On the contrary, a rigid personality efficiently performs optimal solutions of previously practiced situations using previously fixed mental and behavioral sets, while displaying decisiveness and predictability. On one hand, extreme plasticity develops an inconsistent and undependable personality. On the other hand, extreme rigidity leads to stubborn behavior, which perceives ambiguous situations as 'threats' (*21*). Thus, optimal levels of creativity require cycling plasticity- and rigidity-dominated mindsets.

The "blind-variation and selective retention model of creativity" (*89*) is, in fact, describing the same plasticity-rigidity cycles that were detailed before as a major mechanism of animal and human learning. Importantly, brainstorming involves separated plastic (idea-generating) and rigid (idea-selecting, idea-combining) segments (*90*). Last but not least, Kuhn's concept on scientific revolution (*91*) also describes long periods of relative rigidity interrupted by short segments of high plasticity in conceptual continuity. Thus scientific progress also resembles a large-scale plasticity-rigidity cycle.

The social aspects of creativity (like community-aided idea selection) were emphasized by Csíkszentmihályi (*92*). In the network context creativity is often displayed by special, creative nodes dynamically bridging a large number of distant network segments (*45*). However, excess individual creativity can be detrimental to society, because creators invest in their unproven ideas at the expense of propagating proven ones (*93*). Excess creativity is related to the "price of anarchy" in game theory (*94*) showing the degradation of system's efficiency due to the selfish behavior of its agents. Importantly, many individuals can benefit from the creativity of the few without being creative themselves by copying creators (*36*). Finding the optimal level of group-creativity may require plasticity-rigidity cycles of social groups. In agreement with this assumption, an intermediate amount of long-range connections (*35*) resulting in the simultaneous presence of boundary spanning brokerage and trust-building closure, as well as rotating leadership and contribution (*37*) were shown to be key factors of team-success in business, arts, sports and science. Changes of group-plasticity and rigidity dominance may be an important mechanism of plasticity-rigidity cycles of social groups detailed in the next section.

***Adaptation of social groups.*** The organizational learning cycle (*95*) has the same two major phases of plastic exploration and rigid selection/action, like the creative thinking process described above. The exploration/exploitation trade-off [i.e. the two phases of organizational learning exploring new possibilities and exploiting existing certainties (*24*)] resembles to the plastic-rigid duality again. Changing dominance from exploration to exploitation was shown to be useful in early and late phases of firm and product development as well as in plastic and rigid business environments (*96,97*). Task switching (*98*), the PDCA-cycle [plan-do-check-act/Shewhart/Deming-cycle (*99,100*)] and the OODA-loop (*101*) are all plasticity-rigidity cycle variants helping decision making and process control. All these examples describe plasticity-rigidity cycles at the level of individuals and their social groups. In light of these findings, the controversy, whether the business growing strategy of diversification or focusing is better (*102*), may be resolved by their separate, alternating use, which would correspond to a plasticity-rigidity business learning cycle.

Alternating dominance of exploration and exploitation may be observed on levels of national or global economies. Schumpeter's business cycle theory (*103*) described multiple and overlapping scales of economic cycles having innovative/expansive and stagnating/selective periods (Fig. 2D), which resemble to plastic and rigid phases of plasticity-rigidity cycles, respectively. It will be a rather interesting subject of further studies to clarify, whether these and other business cycles actually correspond to adaptation mechanisms – at the scale of macro-economy.

Globalization of the market encouraged system-level approaches to improve the structural stability of macro-economy segments, such as that of banking ecosystems. As the 2008 financial crisis showed, excessive plasticity of the banking ecosystem (displayed by excessive system homogeneity or by excessive dimensionality of derivatives) can minimize the risk for each individual bank, but also maximizes the probability of the entire system collapsing (*104,105*). The lessons of the 2008 crisis led to counter-cyclic regulatory protocols making the banking ecosystem more rigid in booms and more plastic in recessions. These regulatory measures may increase capital and liquidity requirements, bank balance sheet, business model and risk management diversity, as well as banking ecosystem core dominance, hierarchy and modularity in booms, and do the reverse in recessions. I will describe other lessons of plasticity-rigidity cycle-based intervention design later.

*Adaptation of ecosystems*. Ecosystems show a remarkable structural stability, which was perturbed much less by numerous glacial-interglacial transitions in 400,000 years than by human interventions of the last two centuries (*106*). Structural stability is increased by the emergence of ecosystem communities, and by ecosystem nestedness (*23,29,63*). Nestedness marks a core-periphery network structure, where generalist species form the core interacting with other generalists and specialists, while specialists form the periphery interacting only with generalists. Structurally stable ecosystems are more stable against invasions by new species, or environmental changes, but may return more slowly to their equilibrium position after perturbation than structurally less stable ecosystems (*29*). Ecosystem networks have a remarkably conserved structure from the Cambrian Period to modern times (*107*).

Ecological history shapes ecosystems' structural stability (*108*). The importance of ecosystem adaptation-history resembles to the metaplasticity of neuronal networks (*17*) described before. Importantly, the Cambrian explosion produced less hierarchical food-web network structures than those observed in modern times (*107*), which may imply larger ecosystem plasticity in resource-rich conditions. Phenotypic plasticity of invaders increases their invasion success (*108*). Phenotypic plasticity of resident species also increased their ability to oppose invaders. Importantly, system plasticity increased the steepness of the fitness landscape for the invader, making the invasion more difficult even for phenotypically plastic invaders (*109*). This latter study shows a beautiful example, where plasticity of the system nodes increases system plasticity, which leads to a larger structural stability of the system making it less vulnerable against invasions.

From these latter examples the general view emerges, that pulse-like changes in ecosystem environment induce the appearance/survival of more plastic ecosystems, which may return to a more rigid structure after the environment has been (re)stabilized again. Thus, plasticity-rigidity cycles of ecosystems may operate in long, lifetime-spanning, or quasi-evolutionary timescales. I will detail other key examples of long-term consequences of plasticity-rigidity cycles in the next section.

## Long-term consequences and applications of plasticity-rigidity cycles

I listed several examples and applications of plasticity-rigidity cycles in the previous section. Their vast majority described short-term adaptation processes. In this section plasticity-rigidity cycles of whole-life or evolutionary time-scales are described. I will also summarize the possibilities to use the modifications of plasticity-rigidity cycles as beneficial therapeutic interventions in medicine and management of social or ecosystem crises.

***Aging as a diminished capacity of plasticity-rigidity cycling.*** The first example of long-term adaptive changes is aging. Aging is increasingly considered as a complex network phenomenon. An aged organism has already collected an aggregated impact of adaptation over the lifespan with a summarized effect of tradeoffs, sub-optimal resource distributions, compromises and collateral damage, also called as allostatic load. Aging can be regarded as a price we pay to achieve and maintain complexity by the cooperation of self-organizing sub-systems (*4,110*).

The cumulative effects of life-long adaptive processes induce an increased rigidity of aged organisms. Cognitive functions show an increasing rigidity with aging. Performance in psychological tests of fluid intelligence tends to decrease with age. This is in sharp contrast by performance improvements in situations, which were already practiced. Rigidity of the personality decreases between ages 5 and 18, remains fairly stable up to age 60 and increases afterwards. Low scores of mid-life psychological rigidity were found to predict high levels of intellectual functioning in old age (*21*).

Recently Zhou et al (*28*) suggested that epigenetic modifications may make genetic regulatory networks more rigid by increasing system constraints. Comparing this assumption with the recent finding that DNA methylation status is an excellent predictor of the age of human tissues and cells (*111*), gives yet another suggestive clue that increased system rigidity may be a key determinant of the aging process. Indeed, age-induced cognitive decline is associated with an epigenetically-mediated decrease in synaptic plasticity, where histone acetylation plays a key role (*112*). Histone- and DNA-mediated epigenetic lock of plasticity-rigidity cycles in their rigid phase may induce an age-dependent deterioration of the adaptation potential. This interesting hypothesis waits for experimental testing.

***Plasticity-rigidity cycles in evolution.*** There are many similarities between major concepts of evolution and plasticity-rigidity cycles. Evolution proceeds *via* combined variation and selection steps. These two major tenets of evolution resemble the exploration/discovery-type and optimization/selection-type steps of plasticity-rigidity cycle-like optimization processes. Evolvability [ability to generate phenotypic variation) and canalization (ability to maintain the same phenotype (*18*)] may serve as representations of plastic and rigid system behavior, implying many or few system responses, respectively. Finally, punctuated equilibrium [as cycles of neutral diversity expansion followed by selective diversity contraction (*113*)] may be regarded as serial dominance-shifts of plastic and rigid behaviors. I am aware that far-fetched analogies pave a dangerous path. This is especially true, if thinking on the multiple dimensions of interactions between complex organisms, their populations and environment including conscious, learned responses (*114*) and multi-generation timescales. However, I will list a number of examples, which make the assumption that "*plasticity-rigidity cycles of various time-scales and ranges help evolution*", more plausible.

The first two examples are rather indirect. They show that 1.) moderate environmental variations (changing both the optimum of system response and selection pressure) induce better capability for further adaptation than constant environments (*115*); and 2.) organisms displaying intermediate levels of plasticity have a faster adaptation speed than either too plastic or too rigid counterparts [if their attractor structure is complex and/or changing, thus finding its optimum is not trivial (*25*)]. The beneficial effects of environmental changes and the necessity to find an optimal structural stability may both reflect and require the help of plasticity-rigidity cycles.

The next two examples describe evolution of model and *in vitro* systems, where alternating plasticity- and rigidity-dominated steps were needed for optimal adaptation. Plasticity (epigenetic variation) and memory (efficiency of inheritance increasing selection strength) enhanced each other's effect to steer a model organism towards optimal adaptation.

Consecutive increase and decrease of system's plasticity was required for passing convex and concave segments of the epigenetic landscape, respectively, in the path towards the optimum (*116*). Similarly, SELEX-type in vitro evolution processes involving serial selection and enrichment steps converged to their optimum better, if mutation was allowed before the selection steps (*117*). Introduction of mutations increased the similarity of these *in vitro* evolution experiments to plasticity-rigidity cycle-like optimization processes further.

The last and most convincing examples demonstrate that long periods of exploration/discovery at the genotype level often precede sudden bursts at the phenotype level as shown in the *in vitro* evolution of tRNA (*118*), 3000 generations of laboratory *E. coli* evolution (*119*), as well as by the *in vivo* evolution of HIV-1 (*120*) and H3N2 influenza viruses (*121*). Such bursts of the phenotype are prepared and helped by the cryptic accumulation of genetic variation (*113*), whose phenotypic appearance is mediated by the medium-effects of chaperones, prions and network structure (*3,49-51*) as described earlier. Recent examples indicated that accumulation of cryptic changes followed by sudden bursts of system behavior may be a property of the onset of several diseases (*26,31,122*), as well as critical transitions of ecosystems, markets or climate (*38*). Importantly, stabilizing selection decreases phenotypic plasticity again (*123*). These findings suggest that alternating exploration and selection steps may be a general pattern of complex system behavior.

In conclusion, [citing Andreas Wagner (*113*)] "*evolutionary adaptation proceeds by cycles of exploration of a neutral network, and dramatic diversity reduction as beneficial mutations discover new phenotypes residing on new neutral networks.*" Here neutral network exploration represents a gradual increase in system plasticity, while the diversity reduction after a sweep of beneficial mutation represents a transient increase of system rigidity. The numerical calculation of network plasticity/rigidity in various stages of evolutionary processes will be an exciting task of future studies.

***Intervention design: medical therapies and crisis management.*** Plasticity-rigidity cycles offer a wide range of novel possibilities for efficient system control and intervention design. Allowing or accelerating plasticity-rigidity cycles may help complex systems to find their global optimum, i.e. largest structural stability. Activation of plasticity-rigidity cycles by 'noise-enhancer drugs' may activate latent parasites, such as HIV viruses, which allows their complete eradication (*46*). Conversely, 'freezing' of plasticity-rigidity cycles may prevent the rapid evolution of harmful systems, such as cancer stem cells, and may allow a targeted action on their either plastic or rigid forms (*54*). Introduction optimally designed plasticity-rigidity cycles to education (e.g. in form of learning/unlearning/re-learning) or to psychology may lead to improved results (*17*).

Plastic and rigid systems should be targeted in a different fashion requiring 'central hit' or 'network-influence' strategies, respectively (*31*). Moreover, network plasticity may differ from patient to patient (*5*) and changes with biological age (*4*), which require network plasticity-graded personalized therapies. Disease development often proceeds *via* rigidity → plasticity → rigidity transitions (*26,31,122*). Importantly, both plasticity and rigidity transitions offer 'windows of opportunity', where the system can be easily re-programmed approaching or passing a bifurcation *via* a critical transition (*38*). Recently high-capacity methods were constructed to find intervention target node sets [(*31*) and Szalay & Csermely, in preparation], which give excellent tools for efficient intervention design.

Though most of the examples so far were taken from the field of medicine, the same principles can be used for finding efficient interventions in the development of new materials, increasing teaching efficiency, group productivity and business growth (Table S3 of Supplementary Information). Plasticity-rigidity cycle-dependent targeting may also be implied in efficient financial regulation or in the preservation of ecosystems (*104,105*) providing novel options for complex system crisis management.

**Conclusions and perspectives**
In conclusion, I showed by i) thorough characterization of plastic and rigid states and their relationship to the system's structural stability (Boxes 1 and 2, Fig. 1, Table S1 of Supplementary Information); ii) several possible mechanisms (Table S2 of Supplementary Information); and iii) a number of salient examples (Fig. 2, Table S3 of Supplementary Information), that *plasticity-rigidity cycles constitute a powerful, general, system-level adaptive mechanism* helping the development of structurally stable complex systems.

Though structurally stable states can be accessed by relatively few changes of suboptimal systems (*30*), finding them in the vast parameter space still requires special mechanisms in both constant and changing environments. On one hand, if the environment is stable for a long time, long-term adaptation leads to the development of highly specialized, rigid structures. Finding of these highly optimized states requires the help of plasticity-rigidity cycles, since otherwise the system would easily be trapped in a local optimum of the rough state space. On the other hand, if the environment is fluctuating, escape from local minima becomes much easier, but the exact configuration having a high structural stability keeps changing. To find this always changing, elusive stability condition the system needs a constant maneuvering, where plasticity-rigidity cycles again offer a great help. Thus either the aim is clear, but the path leading there becomes difficult, or the path is rather smooth, but the aim becomes undefined. In both cases, as shown by the key examples of assisted protein folding, cell differentiation, learning, as well as establishing financial and ecosystem structural stability (Fig. 2, Table S3 of Supplementary Information), plasticity-rigidity cycles offer an efficient way of adaptation. Plasticity-rigidity cycles, in fact, utilize and extend the duality of Archilochus' famous saying that "*The fox knows many things; the hedgehog one big thing*" elaborated by Isaiah Berlin (*124*).

I share the worries of Simon that "*our speculations have carried us over a rather alarming array of topics, but that is the price we must pay if we wish to seek properties common to many sorts of complex systems*" (*1*). Plasticity-rigidity cycles are obviously not the only successful adaptive mechanisms. There are many other ways to restore the system's original status or explore alternative options (*4*). As another limitation of the concept of plasticity-rigidity cycles, a general theory for the numerical characterization of system plasticity and rigidity has not been developed yet (*32*). Importantly, on higher levels of system complexity of social and ecosystems, many simultaneous plasticity-rigidity cycles may operate in different time-scales and subnetworks, like that demonstrated by Schumpeter's business cycles [Fig. 2D; (*103*)]. Their multiple overlaps may prevent the observation of cyclic behavior. Additionally, some of the examples (such as those on ecosystems and evolutionary processes) were only indirect. Despite of these limitations, the widespread occurrence and robustness of plasticity-rigidity cycles gives great hopes that the further applications of plasticity-rigidity cycle-based optimization, as well as the plasticity-rigidity cycle-based intervention design methods that were listed in the previous section, will be increasingly important and successful.

I hope that this work will prompt to develop more general numerical measures of structural and functional plasticity and rigidity. Examination of plasticity-rigidity cycles i) displayed by interacting systems; ii) observed during environmental changes; iii) working in node- and edge-structures of multilayer networks of cells, brain, society or ecosystems or iv) operating during evolution will also be exciting tasks of future studies. In an early study Dunham *et al.* (*125*) described iterative improvement as the natural framework for heuristic design. Here I demonstrated that plasticity-rigidity cycles form a natural framework for adaptation in all areas of organization and life.

## ACKNOWLEDGMENTS

The original idea of this paper was conceived during a residence at the Rockefeller Foundation Bellagio Center. the author apologizes for those more than 150 respected colleagues, whose key contributions could not be cited in this paper due to length limitations. The help of A. Kovács in designing Fig. 1B is acknowledged. I thank members of the LINK-group (www.linkgroup.hu) of the Department of Medical Chemistry at the Semmelweis University, Budapest, Hungary for helpful suggestions. The useful comments of M. Gáspár (including the use of transition probability distribution entropy as a potential plasticity/rigidity measure) are gratefully acknowledged. Comments of M. Antal, D. Gyurkó, G. Kalmár, T. Korcsmáros, M. Kormos, K. Szalay and A. Szilágyi (LINK-Group), as well as B. Aczél, G. Baffy, T. Freund, B. Hangya, G. Klein, H. Malmgren, B. Papp, G. Papp, C. Sőti, A. Steták, L. Székely and G. Székely are especially appreciated. Special thanks for the encouragement and help of H. J. M. Kiss. Work in the author's laboratory was supported by a research grant from the Hungarian National Science Foundation (OTKA-K83314). See the supplementary materials for additional references.

## SUPPLEMENTARY MATERIAL

Tables S1 to S3
References

| **Box 1** | **Glossary** |

In this glossary key terms of this paper are defined. Since the concept of plasticity-rigidity cycles spans a wide range of disciplines, the definitions listed here do not necessarily agree with all definitions. This is especially true for terms marked with asterisks, which are used in different disciplines with a different content. *[Note that sentences in brackets are not parts of the definition, but relate the term to other terms in this Glossary.]*

**Adaptation:** Adaptation is the process how the structure (and behavior) of a system changes to become more suited to its changing environment.

**Attractor:** system's state towards which a system tends to change, regardless of the starting conditions of the system. *[Simple attractors can be perceived as 'basins', if the system's state space is restricted to the 2 dimensions of the x-y plane, and the z-axis denotes an energy-type measure as shown on Fig. 1B. For the sake of simplicity, I do not consider cycle, torus or strange attractors, when mentioning the term attractor in the text.]*

**Canalization:** Canalization is the ability to produce the same phenotype regardless of environmental or genotype variability.

**Complex/complexity*:** Complex systems display emergent properties (i.e. properties, which can not be described or predicted from the properties of their components; e.g. from the properties of the nodes of complex networks). The definition of numerical measures of complexity proved to be notoriously difficult (*1,3,4,23*).

**Core:** the network core is a central and densely connected set of network nodes. *[The network periphery has sparsely connected, usually non-central nodes, which are preferentially connected to the core (31). The development of network core increases the structural stability of ecosystems (29,63).]*

**Evolvability:** Evolvability is an organism's ability to generate heritable phenotypic variation (*18*).

**Flexibility*:** flexibility characterizes largely reversible and rapid responses of (partially) rigid systems, often returning the system to its former equilibrium. *[Flexibility is often used as a quasi-synonym for plasticity. Since the above properties imply that flexibility is often a property of (partially) rigid systems, I prefer to use the term plasticity instead of flexibility in this paper.]*

**Hub:** hubs are highly connected nodes of a network. Usually a hub has more than 1% of total connections.

**Learning*:** learning is perceived as a process of acquiring and encoding novel information, which modifies existing information content, and produces permanent changes of the system. *[Learning increases system rigidity, which may include both a specific increase of rigidity at selected points (memory-consolidation) and the deletion of aspecific memory-traces leading to an increase of relative rigidity of the remaining memory content (17).]*

**Memory*:** the term memory is used as an ability of a complex system to store information. *[This definition does not involve the existence or change of previously stored information, and incorporates neither memory encoding/retrieval processes, nor declarative, nondeclarative, short-, intermediate- and long-term memory types (17). The definition of memory is used here as a constraint, and is related to the system's ability to increase its rigidity. Note that complex memory-related processes, like memory retrieval, may decrease with increased brain rigidity.]*

**Module:** network modules (or communities) are groups of network nodes that are relatively isolated from the rest of the network, and are connected (and most of the time functionally linked) to each other.

**Network:** a network is a set of nodes, which are connected with network edges to each other.

| **Box 1** | **Glossary** (continuation) |
|---|---|

**Node:** nodes are the building blocks of a network. Nodes are also called as network elements, vertices in graph theory, or actors in sociology. *[Nodes of most real networks are not abstract points (like vertices in mathematics), but complex (sub-)networks themselves.]*

**Noise*:** noise reduces the accuracy of signal transduction (where 'signals' were learned and stored in the system's memory earlier as preferred ways of communication). *[Noise often emerges from stochastic processes, increases system plasticity, and often reflects the plasticity of the sub-networks of system constituents (network nodes).]*

**Plasticity*:** the term of plasticity is used as functional plasticity meaning that the system has a large number of possible responses to (even) unexpected environmental changes. *[Plastic systems have a high capacity of adaptive change, and easily display a different phenotype/behavior developing a larger diversity of plastic systems in response to varying environmental conditions. Networks describing plastic systems display signs of structural plasticity.]*

**Rigidity*:** the term rigidity is used as functional rigidity meaning that the system has a few possible responses to an external influence (*32*). *[Structural rigidity in a strict sense is a mathematical theory describing the degrees of freedom of ensembles of rigid physical objects connected together by flexible hinges. The term of structural rigidity is used to describe the network topology of functionally rigid systems.]*

**Robustness*:** the persistence of a system's characteristic behavior (even) after unexpected environmental changes (*2*). *[System robustness helps to achieve high structural stability. Robustness is high, if the system has both plastic and rigid properties.]*

**Stability*:** in this paper stability denotes structural stability (*22*), which is high, if 'smooth' (continuously differentiable) perturbations preserve the homeomorphism of the changing attractor structure of the system. *[Topologically non-equivalent attractor structure develops, if the system passes a bifurcation. Structural stability can be characterized by the range of conditions, where the system shows qualitatively the same responses (29). In this paper the concept of structural stability over Lyapunov stability or stochastic stabilities (126) is preferred (23). The two types of stabilities do not necessarily change in parallel (29). Lyapunov and stochastic stabilities may fluctuate during plasticity-rigidity cycles (decreasing in plastic states, while increasing in rigid states). High structural stability requires both plastic and rigid system properties, and may increase as plasticity-rigidity cycles proceed.]*

**System*:** A group of interacting, interrelated, or interdependent parts forming a complex whole. *[Systems are often relatively separated from their environment, and display a high level of independence. However, what is a 'system' for one process, may be the 'environment' of another and vice versa. Note the similarities between the definition of a module and a system. Though the two definitions do overlap, systems are usually more independent than modules (3).]*

| **Box 2** | **Numerical descriptions of system plasticity and rigidity** |

Plasticity and rigidity refer to high and low system capacity of adaptive change, respectively (Box 1). Though plasticity has been extensively characterized in material science numerically (*16*), the system-level generalization of these studies is missing. The mathematical theory of structural rigidity was born by the 1864 paper of Maxwell (*19*) on constraint counting. Structural rigidity can be measured by a number of combinatorial graph theory approaches (*32*). The widely used pebble game algorithm performs rigid cluster decomposition determining redundant edges and the system's internal degree of freedom. This algorithm was successfully applied to characterize the structural rigidity of protein structures [http://kinari.cs.umass.edu; (*68*)]. Recently, a generalized pebble game algorithm was developed using non-integer probabilistic constraints (*127*). The extension of the Laman theorem (*20*) describing structural rigidity conditions of rod/joint networks in 2 dimensions to 3 dimensions is a famous open problem of mathematics. The determination of independent system constraints proved to be especially difficult. Matroid theory provides promising approaches to solve this latter problem (*32*). Mesoscopic extension of local and global network entropy measures (*6*) may provide promising approximations of structural plasticity and rigidity. Numerical measures of functional plasticity/rigidity (defined as the number of possible system responses, which is related to the number of possible changes between system attractors; see Box 1) may depend on the type of perturbation. However, structurally stable systems have a stable attractor structure, where the number of system responses becomes more independent from the type of smooth (continuously differentiable) perturbations. In probabilistic systems functional plasticity/rigidity can be approximated by the entropy of the system's transition probability distribution. Importantly, a general theory linking structural and functional rigidities has not been developed yet (*32*). The recent concepts of information topology (*128*) or ruggedness (*129*) of attractor landscape may open new approaches. In conclusion, the system-level quantification of structural and functional plasticities and rigidities, similarly to that of other emergent properties of complex systems, such as robustness or complexity itself, proved to be notoriously difficult. I hope that this work will encourage the development of more general numerical measures of structural and (most importantly) functional plasticity and rigidity.

Figure 1

**Fig. 1. Plasticity and rigidity as two major states of complex systems.** **(A)** Plasticity and rigidity emerge as twin-hubs of a rich, multi-disciplinary conceptual framework. The central circular arrow-pair of the panel refers to plasticity-rigidity cycles. Most of the concepts of the panel are defined in Box 1. Lines highlight a few major connections between the concepts listed on the panel. Yellow, blue and green boxes denote concepts more (but not exclusively) related to plasticity, rigidity or both, respectively. **(B)** Illustrations of possible state space representations of plastic and rigid systems. The four figures show state spaces of plastic (left side) and rigid systems (right side), where the horizontal plane represents the possible states of the system, while the vertical axis illustrates an energy-type measure. Plastic and rigid systems have smooth and rough state spaces, respectively. The top two figures illustrate the state space of a rather simple plastic (left) and rigid (right) system having a single attractor. While the plastic system displays a low Lyapunov stability returning to its attractor slowly, the rigid system has a high Lyapunov stability showing a fast return to its attractor (note that Lyapunov stability can be extended to stochastic systems, see Box 1). The bottom two figures illustrate the state space of more complex systems with multiple attractors. Each of the attractor basins of these complex systems displays a different Lyapunov stability. In a changing environment neither extremely plastic nor extremely rigid systems have high structural stability (*22*), where structural stability refers to the situation, when 'smooth' (continuously differentiable) perturbations do not induce qualitative changes in the system's attractor structure. Extremely plastic, 'fluid' systems shift their attractor structure easily. Extremely rigid systems are fragile, which leads to their decreased structural stability inducing large gross changes in their attractor structure, once the environmental changes go beyond the limited response-set acquired and refined by the rigid systems during their past experiences. However, if the environment remains constant for a long time, a rigid system may possess a highly optimized, efficient response. In a changing environment high structural stability is characteristic to complex systems displaying the signs of both plasticity and rigidity. Importantly, a rigid system trapped in a local minimum (shown by the upwards arrow) may increase its plasticity (bottom left), and acquire a similar, but different attractor structure after increasing its rigidity again (bottom right). Such changes in system plasticity and rigidity may be repeated several times helping to find the system's global optimum and maximal structural stability. I term this adaptation mechanism as plasticity-rigidity cycles, and show its generality by the examples of this paper including those of Fig. 2.

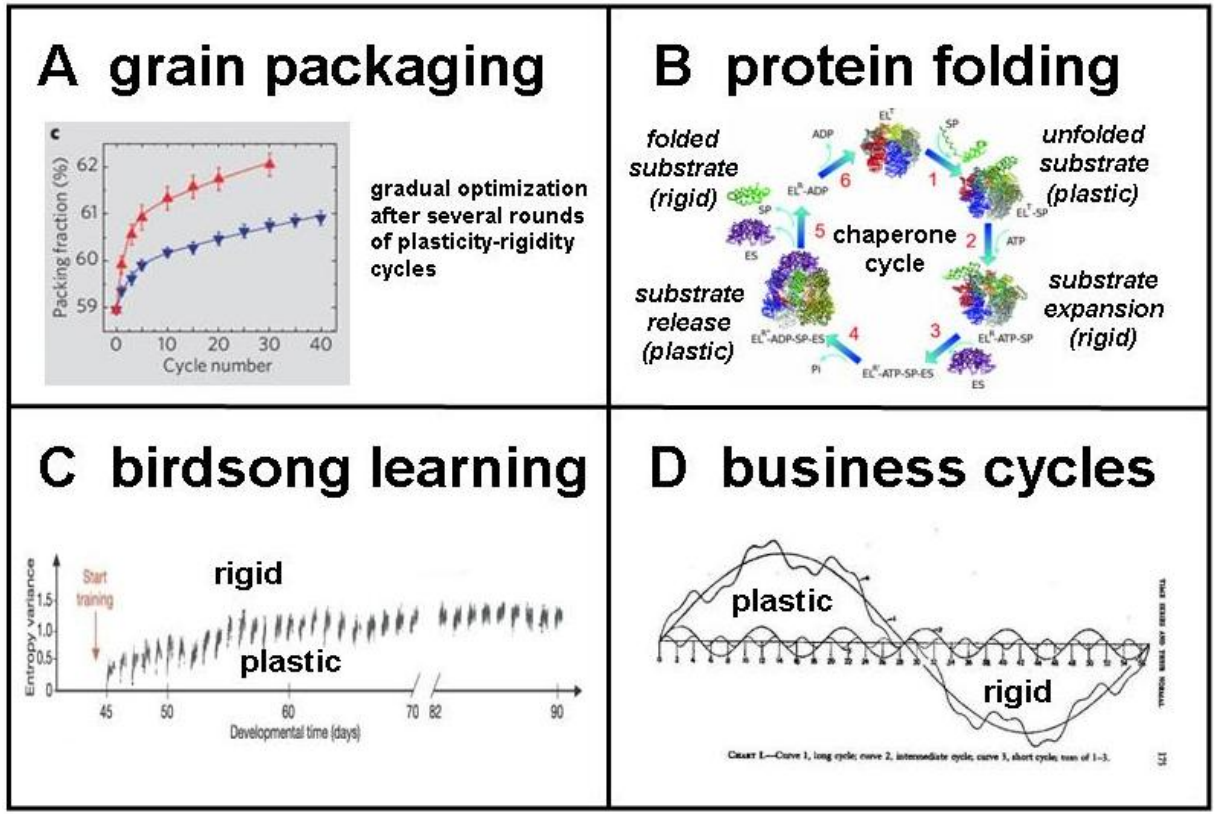

**Fig. 2. Key examples of plasticity-rigidity cycle-mediated adaptation mechanisms.** The figure illustrates plasticity-rigidity cycles by four examples covering a wide range of disciplines and system complexity. **(A)** Increase of packaging efficiency after thermally-induced plasticity-rigidity cycles of glass spheres in a plastic cylinder [reproduced with permissions from Chen et al. (*41*)]. **(B)** Molecular chaperones help protein folding and protein complex assembly *via* multiple, ATP hydrolysis-driven cycles inducing the extension and consequent release of substrates (*43*) inducing their larger rigidity and plasticity, respectively, illustrated by the chaperone cycle of the GroE bacterial chaperone [reproduced with permissions from Stan et al. (*42*), Copyright (2007) National Academy of Sciences, U.S.A.]. **(C)** Cyclic development of song beauty (represented as Wiener entropy variance) in a 45-day training period of zebra finches. Birds that showed larger cycling in their song complexity learned better [reproduced with permissions from Derégnaoucourt et al. (*40*)]. Later studies extended these findings to other bird species and to human infants (*79*), showing that neuronal plasticity-rigidity cycles of the HVC bird song nucleus accompanied the sequential birdsong learning steps (*80*). **(D)** Business cycles described by Schumpeter (*103*) have several overlapping innovative/expansive phases and stagnating/selective phases. Business cycles increase overall productivity, and can be regarded as an adaptation process at the level of macro-economy. [Reproduced with permission from Schumpeter (*103*).] Organizational learning cycle (*95*) and changes of exploration and exploitation phases (*24*) describe other economy-related examples, where the operation of plasticity-rigidity cycles was more directly demonstrated.

Supplementary Materials for

# Plasticity-rigidity cycles:
# A general adaptation mechanism

Peter Csermely*

*Department of Medical Chemistry, Semmelweis University, Budapest, Hungary*

*Corresponding author. E-mail: csermely.peter@med.semmelweis-univ.hu

**This PDF file includes:**

**Table S1. Properties and detection possibilities of plastic and rigid systems**

| Property | Plastic system | Rigid system | Remarks and limitations |
|---|---|---|---|
| *Properties related to functional rigidity (in this paper: rigidity)* | | | |
| system's temperature | high | low | temperature here is not physical temperature in most cases, but used in a transferred sense as in simulated annealing (*1-3*) |
| state of system | liquid | solid | the words 'liquid' (fluid) and 'solid' (condensed) are used here only as analogies |
| transmission of perturbations | low | high | this note refers to the transmission of perturbations (in this sense: signals) with a relatively preserved information content, i.e. with low dissipation; the terms 'low' and 'high' describe only an average behavior where exceptions may occur (*4*) |
| dissipation of perturbations | high | low | dissipation is used here as a measure of diminished perturbations during the transmission process; there might be exceptions from the general properties described by 'high' and 'low' (*4*) |
| population of system's nodes or multiple systems | diverse | uniform | this difference (as most of the others) is only relative, thus populations of rigid system nodes or systems are not fully uniform in most cases |
| population dynamics of system's nodes or multiple systems | variation | selection | variation and selection are meant here in their Darwinian (*5*) sense, where both the system and its nodes (constituents) are living organisms; plastic and rigid systems may display both properties [see section on evolutionary aspects for more details (*6*)] |
| duration of constant system properties | short | long | plastic states may last long, and rigid states may have a very short duration; however, plastic systems often drift, while a stable (in the sense of unchanged, 'frozen', or 'dormant') state is more characteristic to rigid systems |
| duration of system responses to a usual, previously occurred environmental change | relatively slow, if compared to that of rigid systems | fast | in nature 'usual, previously occurred' or 'novel, unexpected' environmental changes do not exist in their extreme forms described in this Table; responses of flexible (see Box 1 for differences in definition of flexibility and plasticity) systems to a usual, previously occurred environmental change can be very fast; changes in evolutionary timescale often follow a punctuated equilibrium (*7-22*) |
| duration of system responses to a novel, unexpected environmental change | relatively fast, if compared to that of rigid systems | can be very slow and, if occurs at all, unexpected (e.g. the rigid system breaks) | |
| state space of the system | smooth | rough | in a rigid system there is a limited number of transitions between various attractors, while a plastic system may easily 'drift' between its attractors significantly increasing the number of possible transitions (*9,11,23-30*) |
| *Properties related to both functional and structural rigidities* | | | |
| system's resources | high | low | resources here may refer to the system's energy level, but are primarily used as a measure of the system's ability to build and maintain connections between its nodes |
| information encoded by the system | low | high | information is meant here as information encoded by the system using Shannon's (*31*) term, which does not necessarily equal with the 'retrievable' information from the system |

**Table S1. Properties and detection possibilities of plastic and rigid systems (continuation)**

| Property | Plastic system | Rigid system | Remarks and limitations |
|---|---|---|---|
| *Properties related to structural rigidity* | | | |
| system's entropy | high | low | entropy here refers to the entropy of the network (*32-35*) describing the complex system; note that this description of entropy is related to information theory and the degrees of freedom of networks embedded in a physical space (see Box 2 for the connection of rigidity) |
| connections between system network nodes | weak, sparse | strong, dense | the number and weight of connections refer to an average |
| dominant mode of system (and network) connection dynamics | disassembly | assembly | in most systems assembly and disassembly processes occur simultaneously; disassembly often co-occurs with differentiation, while assembly with integration (*36*) |

For definitions and measurement of the emergent system properties (*37,38*) of stability (*39-43*), plasticity, as well as of structural and functional rigidities (*44-48*) see Boxes 1 and 2 of the main text. Note that plastic and rigid forms often co-exist in complex systems (or in populations), and often change to each other displaying a high level of dynamics. Note that most statements of this table are only hypothetical in their general form described here, and the statements on functional and structural rigidities may overlap.

**Table S2. Possible mechanisms inducing plasticity-rigidity cycles**

| Possible mechanism | Role in the development of plasticity-rigidity cycles |
|---|---|
| *General, network-independent mechanisms* | |
| changes of environment- or node-derived (extrinsic or intrinsic) noise | noise-fluctuations (such as those of pink or 1/f noise) may induce plasticity-rigidity cycles (*28,49-55*) |
| changes in the 'viscosity' of the 'medium' surrounding system nodes | water-mediated plasticity fluctuations in assisted protein folding (*56*) and cellular function\*; changes in membrane fluidity\*; chaperone-mediated, stress-dependent changes of phenotypic plasticity (*57*); neuromodulation by e.g. volume transmission changing synaptic plasticity (*58-60*); changes in the available money (*61*), innovation and/or trust (*62*) inducing plastic surplus and rigid shortage economies like (in extreme forms) capitalism and socialism (*63*) |
| *Network structure-dependent mechanisms* | |
| changes of internal nodal noise, plasticity and rigidity; propagating effects of soft spot and rigidity seed nodes | nodes of most real world networks are networks themselves: their changing noise and plasticity/rigidity may induce softening or rigidity transitions at the level of their whole system, or at the level of separate system segments, such as the network core or modules (*64-67*) |
| decreased and increased hub (especially party hub) dominance | decreased system resources or increased environmental stress induces the emergence of hubs followed by the appearance of mega-hubs, stars (*68-70*); plastic and rigid phases may display decreased or increased hub-dominance\* |
| alternating large/fuzzy or small/compact network core, rich club, bow-tie and nested network structures | large/fuzzy network cores increase system degeneracy (*71*) and robustness, while small/compact cores make the system highly controllable and induce non-reversibility (*70,72-76*); core-size alternation (including increased and decreased levels of hub-repulsion) may be a major mechanism of plasticity-rigidity cycles |
| alternation of fuzzy, overlapping, stratus-type modules and compact, segregated, cumulus-type modules | modules allow the parallel optimization to multiple requirements: overlapping modules may operate in plastic, while compact modules may characterize rigid phases (*4,77-80*); however, plasticity-rigidity cycles may also occur in different modules separately |
| changes in the extent of network hierarchy | plastic and rigid phases of plasticity-rigidity cycles may display higher and lower hierarchies, may be more source- and sink-dominated, and may be characterized by distributed and centralized system control, respectively\* |
| differences of plastic/rigid properties and transitions between network classes, like flow-type and association-type networks | sub-networks of the interdependent, multiplex networks of complex systems may display different levels of plasticity and rigidity at the same time\* (*72*) |

Note that here only those mechanisms are listed, which are general enough to occur on several levels of system organization (such as at the level of molecules, cells and organisms), and omitted the description of several level-specific mechanisms, such as changes in the transcription of specific genes, changes in intracellular transport, etc. For definitions of several terms used in this table see Box 1 of the main text. Asterisks denote mechanisms, which are in part only hypothetical. Recently high-capacity methods were constructed to find intervention target node sets (*4,29,81*).

**Table S3. Examples of plasticity-rigidity cycle-induced adaptation processes and their applications**

| Adaptive system | Description of plasticity-rigidity cycle inducing higher and lower number of possible system responses | Examples of current and possible applications |
|---|---|---|
| molecular assemblies and granular materials | repeated heating and cooling steps induce higher and lower chances of particle rearrangements (*82*) | low-cost increase in tighter packing and alloy/glass homogeneity; development of novel, super-elastic materials |
| macromolecules | molecular chaperones first bind and expand unfolded proteins then release them decreasing and increasing their degrees of freedom, respectively; the consecutive folding of proteins decrease their degrees of freedom again (*83-87*); binding to proteins may proceed *via* repeated changes of plasticity and rigidity (*44-46,48,88*) | increased yield of biosimilar drugs; chaperone and other therapies (*73*) |
| differentiating or dedifferentiating cells | differentiation of progenitor cells or induction of pluripotent stem cells proceeds *via* an intermediate step having more disordered molecular networks and higher cellular heterogeneity showing a higher number of cellular responses than either the starting state or the end state (*89,90*) | juvenilization, regenerative, and anti-cancer therapies |
| neuronal and neural networks, brain | formation of short- and long-term memories, memory consolidation, re-consolidation and retrieval processes proceed *via* cyclic changes of synaptic plasticity and neuronal network dynamics showing cyclic changes of the number of system responses both at molecular and cellular levels (*52,91-111*) | treatments of neurodegenerative diseases and post traumatic stress disorder (*112,113*); increased teaching efficiency (*114*) and fairness in criminal investigations (*115*) |
| conscious mind of individuals and their groups | creativity of individuals and social groups operating *via* dominance-shifts of plastic and rigid mindsets and group structures invoking more and less possible responses, respectively (*116-123*) | increased individual and group productivity, anti-aging strategies (*124-130*), talent support |
| social groups | alternating exploration and exploitation group behavior (*131-134*); business cycles*; boom/recession behavior of the banking ecosystem showing more and less possible system responses, respectively (*43,135,036*) | growth strategies for business firms and economies; counter-cyclic financial regulation (*136-138*) |
| ecosystems | transient increase in ecosystem plasticity in response to an environmental change* enabling a wider range of system rearrangements (*76,139*) | preservation of species diversity and ecosystem structural stability in times of overuse, heavy pollution and global warming (*138,140-142*) |

Series of successful adaptation steps helped the emergence of complexity (*36,71,143-147*). Note that plasticity-rigidity cycles (*148*) are only one of the many forms of system level adaptation mechanisms (*68,70,149-151*). For definitions of several terms used in this table see Box 1 of the main text. The description of plasticity-rigidity cycles and their applications are detailed in sections of the main text "Examples of plasticity-rigidity cycle-induced adaptation" and "Intervention design: medical therapies and crisis management", respectively. Asterisks denote processes or applications, which are, in part, only hypothetical.